\documentclass[aps,prc,twocolumn,superscriptaddress]{revtex4-1}

\usepackage{dcolumn}
\usepackage{graphicx,color,subfigure}
\usepackage{textcomp}
\usepackage{multirow}
\usepackage{relsize}

\begin{document}
  \newcommand {\nc} {\newcommand}
  \nc {\beq} {\begin{eqnarray}}
  \nc {\eeq} {\nonumber \end{eqnarray}}
  \nc {\eeqn}[1] {\label {#1} \end{eqnarray}}
  \nc {\eol} {\nonumber \\}
  \nc {\eoln}[1] {\label {#1} \\}
  \nc {\ve} [1] {\mbox{\boldmath $#1$}}
  \nc {\ves} [1] {\mbox{\boldmath ${\scriptstyle #1}$}}
  \nc {\mrm} [1] {\mathrm{#1}}
  \nc {\half} {\mbox{$\frac{1}{2}$}}
  \nc {\thal} {\mbox{$\frac{3}{2}$}}
  \nc {\fial} {\mbox{$\frac{5}{2}$}}
  \nc {\la} {\mbox{$\langle$}}
  \nc {\ra} {\mbox{$\rangle$}}
  \nc {\etal} {\emph{et al.}}
  \nc {\eq} [1] {(\ref{#1})}
  \nc {\Eq} [1] {Eq.~(\ref{#1})}
  \nc {\Ref} [1] {Ref.~\cite{#1}}
  \nc {\Refc} [2] {Refs.~\cite[#1]{#2}}
  \nc {\Sec} [1] {Sec.~\ref{#1}}
  \nc {\chap} [1] {Chapter~\ref{#1}}
  \nc {\anx} [1] {Appendix~\ref{#1}}
  \nc {\tbl} [1] {Table~\ref{#1}}
  \nc {\fig} [1] {Fig.~\ref{#1}}
  \nc {\ex} [1] {$^{#1}$}
  \nc {\Sch} {Schr\"odinger }
  \nc {\flim} [2] {\mathop{\longrightarrow}\limits_{{#1}\rightarrow{#2}}}
  \nc {\textdegr}{$^{\circ}$}
  \nc {\inred} [1]{\textcolor{red}{#1}}
  \nc {\inblue} [1]{\textcolor{blue}{#1}}
  \nc {\IR} [1]{\textcolor{red}{#1}}
  \nc {\IB} [1]{\textcolor{blue}{#1}}
  \nc{\pderiv}[2]{\cfrac{\partial #1}{\partial #2}}
  \nc{\deriv}[2]{\cfrac{d#1}{d#2}}

\title{Reconciling Coulomb breakup and neutron radiative capture}

\author{P. Capel}
\email[]{pierre.capel@ulb.ac.be}
\affiliation{Physique Nucl\' eaire et Physique Quantique (CP 229), Universit\'e libre de Bruxelles (ULB), B-1050 Brussels}
\affiliation{Institut f\"ur Kernphysik, Technische Universit\"at Darmstadt, 64289 Darmstadt, Germany}

\author{Y. Nollet}
\email[]{yvan.nollet@univ-tlse3.fr}
\affiliation{Physique Nucl\' eaire et Physique Quantique (CP 229), Universit\'e libre de Bruxelles (ULB), B-1050 Brussels}

\date{\today}

\begin{abstract}
The Coulomb-breakup method to extract the cross section for neutron radiative capture at astrophysical energies is analyzed in detail.
In particular, its sensitivity to the description of the neutron-core continuum is ascertained.
We consider the case of $^{14}$C$(n,\gamma)$$^{15}$C for which both the radiative capture at low energy and the Coulomb breakup of $^{15}$C into $^{14}$C$+n$ on Pb at 68~MeV/nucleon have been measured with accuracy.
We confirm the direct proportionality of the cross section for both reactions to the square of the asymptotic normalization constant of $^{15}$C observed by Summers and Nunes [Phys. Rev. C {\bf 78}, 011601 (2008)], but we also show that the $^{14}$C-$n$ continuum plays a significant role in the calculations.
Fortunately, the method proposed by Summers and Nunes can be improved to absorb that continuum dependence.
We show that a more precise radiative-capture cross section can be extracted selecting the breakup data at forward angles and low $^{14}$C-$n$ relative energies.
\end{abstract}

\pacs{24.10.−i, 24.87.+y, 25.60.Gc, 25.60.Tv, 25.70.De}
\keywords{Exotic nuclei, Coulomb breakup, radiative capture, indirect methods in nuclear astrophysics}

\maketitle

\section{\label{intro}Introduction}

In radiative-capture reactions, two nuclei merge to form another nucleus by emitting a photon.
These reactions take place in many astrophysical sites \cite{RR88}.
For example, most of the nuclear reactions that power the Sun are part of the $pp$ chain, which consist mainly of proton captures by light nuclei \cite{BBFH}.
The $s$ and $r$ processes, which take place during explosive stellar transients, like supernov\ae, are sequences of neutron radiative captures \cite{BBFH}.
To provide a precise description of stars, stellar models require reliable cross sections for these reactions.
Unfortunately, the energy range of interest in astrophysical processes is usually quite low, of the order of a few tens of keV, where the cross sections are very small and hence very difficult to measure.
Moreover, many such reactions involve short-lived nuclei, which hinder their direct measurement.
Indirect methods have thus been suggested to bypass direct measurements.

The Coulomb-breakup method, has been suggested by Baur, Bertulani, and Rebel \cite{BBR86}.
Instead of the direct synthesis of a nucleus through radiative capture, this method suggests to measure the dissociation of that nucleus into its more elementary constituents through its interaction with a heavy (high $Z$) target \cite{BBR86,BR96,BHT03}.
Being dominated by the Coulomb interaction, this reaction can be seen as the exchange of virtual photons between the projectile and the target, and hence as the time-reversed reaction of the radiative capture.
At the first order of the perturbation theory its cross section can be related to the radiative-capture one by a simple detailed balance \cite{BBR86,BR96}.
Because experimentalists can then use beams at higher energy and measure reactions with larger cross sections, they should reach higher precision than in direct measurements.

Unfortunately, later theoretical analyses have shown that higher-order effects spoil this nice picture and that a direct extraction of the radiative-capture cross section from Coulomb-breakup measurements is not as simple as expected \cite{EB96,EBS05,CB05}.
Subsequent analyses have then shown that both the breakup of loosely-bound nuclei \cite{CN07} and their synthesis through radiative capture \cite{TBD06} are mostly peripheral, in the sense that they are sensitive only to the tail of the nuclear wave functions.
Relying on these results, Summers and Nunes have suggested a new approach of the Coulomb-breakup method.
Instead of directly inferring the radiative-capture cross section from Coulomb-breakup measurements, they have suggested to extract from the latter an ``experimental'' asymptotic normalization constant (ANC) and use that ANC to compute a reliable radiative-capture cross section \cite{SN08}.

They have tested their idea in the particular case of $^{15}$C, which appears in different astrophysical sites.
The radiative capture $^{14}$C$(n,\gamma)$$^{15}$C is part of the neutron-induced CNO cycle that takes place in the helium-burning shell of light AGB stars \cite{WGS99}.
It also plays a role in the primordial nucleosynthesis of intermediate-mass elements \cite{KMF90}.
Besides its interest in nuclear astrophysics, $^{15}$C is useful to test the Coulomb-breakup method because both its Coulomb breakup into $^{14}$C$+n$ \cite{Nak09} and the radiative capture $^{14}$C$(n,\gamma)$$^{15}$C \cite{Rei08} have been accurately measured.
The ANC Summers and Nunes have obtained by confronting their Coulomb-breakup calculations to the experimental data leads to a radiative-capture cross section in nice agreement with the direct measurements \cite{SN08}.
Their study also confirms that, due to higher-order effects, a fully dynamical breakup model is needed to correctly analyze the breakup reaction, a conclusion also reached by Esbensen in \Ref{Esb09}.

In the present work, we study in more detail the ANC method proposed by Summers and Nunes, focussing on the effect played by the description of the $^{14}$C-$n$ continuum.
In \Ref{CN06}, it was indeed shown that these effects can be significant enough to lead up to 40\% variations in the breakup cross section.
To this aim, we follow Summers and Nunes, and consider the same reaction.
We develop a set of different $^{14}$C-$n$ potentials that produce different descriptions of the $^{15}$C bound state (viz. different ANCs) and of its continuum (viz. different $^{14}$C-$n$ phase shifts).
Using the dynamical eikonal approximation (DEA) \cite{BCG05,GBC06}, we then compute the Coulomb breakup of $^{15}$C on Pb at 68~MeV/nucleon, which corresponds to the experimental conditions of \Ref{Nak09}.
This enables us to analyze the sensitivity of the breakup reaction to the ANC of the $^{15}$C ground state and to its continuum to test the validity of the ANC method developed by Summers and Nunes.

We start this article with a brief presentation of the two-body model of $^{15}$C used in this study, and of the reaction framework we consider (\Sec{theory}).
The $^{14}$C-$n$ potentials developed in this work are provided in \Sec{interactions}.
Our calculations of the Coulomb breakup of $^{15}$C and the radiative capture $^{14}$C$(n,\gamma)$$^{15}$C are presented in \Sec{sensitivity}.
Following the analysis of these calculations, an improvement of the ANC method is suggested in \Sec{improvement}.
The conclusion of this study is drawn in \Sec{conclusion}.

\section{\label{theory}Theoretical framework}
\subsection{\label{nucleus}Two-body description of the nucleus}
We consider two types of reactions involving a nucleus made up of a core $c$, of atomic and mass numbers $Z_c$ and $A_c$, respectively, to which a neutron $n$ is loosely bound: its Coulomb breakup into the core and the valence neutron and its synthesis through the radiative capture of the neutron by the core.
Such a two-cluster system can be described by the Hamiltonian
\beq
H_0=-\frac{\hbar^2}{2\mu}\Delta_{\ve{r}}+V(\ve{r}),
\eeqn{e1}
where $\ve{r}$ is the relative coordinate between the core and the valence neutron and $\mu$ is their reduced mass.
In Hamiltonian $H_0$, the potential $V$ simulates the interaction between the constituents of the nucleus.
It is supposed to be central, but for a possible dependence on the orbital angular momentum (see \Sec{pot}).
In the following, we neglect the spin of the clusters for simplicity.

Within this model, the relative motion of the valence neutron to the core at energy $E$ is described by the eigenstates of $H_0$.
In the partial wave of orbital angular momentum $\ell$ and projection $m$, they read
\beq
H_0\ \phi_{\ell m}(E,\ve{r})=E\ \phi_{\ell m}(E,\ve{r}),
\eeqn{e2}
with
\beq
\phi_{\ell m}(E,\ve{r})=\frac{1}{r} u_{\ell}(E,r) Y_\ell^m(\Omega),
\eeqn{e3}
where the reduced radial wave function $u_\ell$ depends on the norm $r$ of $\ve{r}$ and the angular part $Y_\ell^m$ are spherical harmonics, which depend only on the solid angle $\Omega\equiv(\theta,\varphi)$ \cite{AS70}.

The negative-energy eigenstates of $H_0$ [$E<0$ in \Eq{e2}] correspond to the bound spectrum of the nucleus.
They are discrete and we add the number of nodes $n$ in the radial wave function to the quantum numbers $\ell$ and $m$ to distinguish them.
The reduced radial wave functions of these bound states exhibit the following asymptotic behavior
\beq
u_{n\ell}(E_{n\ell},r)\flim{r}{\infty}{\cal C}_{n\ell}\ W_{0,\ell+1/2}(2\kappa_{n\ell} r),
\eeqn{e4}
where  $\kappa_{n\ell}=\sqrt{2\mu E_{n\ell}/\hbar^2}$, $W$ is the Whittaker function \cite{AS70}, and ${\cal C}_{n\ell}$ is the ANC, which depends on the geometry of the potential $V$ chosen to describe the core-neutron interaction (see \Sec{pot}).

The positive-energy eigenstates of $H_0$ [$E>0$ in \Eq{e2}] describe the continuum of the nucleus, i.e. the states in which the neutron and the core are unbound.
Their reduced radial part behaves asymptotically as
\beq
u_{\ell}(E,r)\flim{r}{\infty}\cos\delta_{\ell}(E)\,j_\ell(kr)+\sin\delta_{\ell}(E)\,n_\ell(kr),
\eeqn{e5}
where $j_\ell$ and $n_\ell$ are the regular and irregular spherical Bessel functions \cite{AS70}, respectively, and the wave number for the neutron-core relative motion $k=\sqrt{2\mu E/\hbar^2}$. The phaseshift $\delta_\ell$ is the only dependence of this asymptotic behavior to the potential $V$.

\subsection{\label{DEA}Breakup model}

The Coulomb breakup of a loosely-bound nucleus corresponds to the dissociation of that nucleus into its more elementary constituents during its collision with a heavy (high-$Z$) target.
This reaction happens because the different constituents of the nucleus do not interact in the same way with the target.
This leads to a tidal force strong enough to break these constituents apart.
On a heavy target, and for a projectile in which one of the two fragments is a neutron, the breakup is mostly due to the Coulomb force.

Using for the projectile the two-body description presented in \Sec{nucleus}, the collision between the projectile $P$ and the target $T$ reduces to a three-body problem.
Within the Jacobi set of coordinates consisting of the relative coordinates between the neutron and the core of the projectile [$\ve{r}\equiv(r,\theta,\varphi)$, see \Eq{e1}] and between the target and the projectile center of mass [$\ve{R}\equiv(R,\Theta,\Phi)$], the three-body Hamiltonian reads
\beq
{\cal H}=-\frac{\hbar^2}{2\mu_{PT}}\Delta_{\ve{R}}+H_0+V_{cT}(R_{cT})+V_{nT}(R_{nT}),
\eeqn{e6}
where $H_0$ is the internal Hamiltonian of the projectile given in \Eq{e1}, $\mu_{PT}$ is the $P$-$T$ reduced mass, and $V_{cT}$ and $V_{nT}$ are optical potentials, which simulate the interaction between the target and each of the projectile constituents.
In \Eq{e6}, $R_{cT}$ (resp. $R_{nT}$) is the relative distance between the core (resp.\ the valence neutron) and the target.

The resolution of this three-body problem corresponds to finding the solution $\Psi$ of the \Sch equation
\beq
{\cal H}\ \Psi(\ve{r},\ve{R})={\cal E}\ \Psi(\ve{r},\ve{R}),
\eeqn{e7}
with the initial condition that the projectile, being in its ground state $\phi_{n_0\ell_0m_0}$, is impinging on the target
\beq
\Psi(\ve{r},\ve{R})\flim{Z}{-\infty}e^{i{\cal K}Z+\cdots}\phi_{n_0\ell_0m_0}(E_{n_0\ell_0},\ve{r}),
\eeqn{e8}
where the $Z$ component of $\ve{R}$ is chosen along the beam axis and the projectile-target relative momentum $\hbar {\cal K}$ is related to the total energy ${\cal E}=\hbar^2 {\cal K}^2/2\mu_{PT}+E_{n_0\ell_0}$.

Various methods have been developed to solve that problem (see \Ref{BC12} for a recent review).
Since the reaction in which we are interested takes place at intermediate energy, we use the Dynamical Eikonal Approximation (DEA) \cite{BCG05,GBC06}.
This reaction model has shown to provide excellent agreement with experimental data for the Coulomb breakup of both one-neutron \cite{GBC06} and one-proton \cite{GCB07} halo nuclei in that energy range.
It properly includes all the couplings within the continuum required to correctly analyze this reaction \cite{SN08,Esb09} and it compares very well with other reaction models \cite{CEN12}.

\subsection{\label{capture}Radiative-capture model}
In the radiative capture, two nuclei merge to form a new nuclide by emitting a photon.
It can therefore be seen as an electromagnetic transition from the continuum of the nucleus to one of its bound states.
The total cross section for the radiative capture from an initial state at energy $E$ in the continuum to the final bound-state $n_0\ell_0$ of energy $E_{n_0\ell_0}$ reads
\beq
\lefteqn{\sigma_{n_0\ell_0}(E)=\frac{64 \pi^4}{4\pi \epsilon_0\hbar v}\sum_{\lambda\sigma}\frac{k_\gamma^{2\lambda+1}}{[(2\lambda+1)!!]^2}\frac{\lambda+1}{\lambda}}\nonumber \\
 &\times &\sum_\ell\frac{2\ell_0+1}{2\ell+1}\left|\langle\phi_{n_0\ell_0}(E_{n_0\ell_0})\|{\cal M}^{\sigma\lambda}\|\phi_{\ell}(E)\rangle\right|^2,
\eeqn{e9}
where $v=\hbar k/\mu$ is the relative velocity between the neutron and the core in the initial continuum state and $\hbar k_\gamma c=E-E_{n_0\ell_0}$ is the photon energy.

In \Eq{e9}, the summation is performed on electric ($\sigma=\rm E$) and magnetic ($\sigma=\rm M$) transitions and on all possible multipoles $\lambda=1, 2,\ldots$
In practice, only a small number of terms are needed to reach convergence.
In the present case, in which we consider a neutron captured by the core, the sole dominant term is $\rm E1$ \cite{TBD06}, for which the transition operator reads
\beq
{\cal M}^{\rm E1}_\mu= e \frac{Z_c}{A_c+1} r Y^{(1)}_\mu(\Omega).
\eeqn{e10}

\section{\label{interactions}Two-body interactions}
\subsection{\label{pot}Different \ex{14}C-$n$ potentials}

With the aim of studying the role of the \ex{15}C description on the breakup calculations and its effect on the extraction of the cross section for the radiative capture \ex{14}C$(n,\gamma)$\ex{15}C, we follow Summers and Nunes \cite{SN08} and develop various potentials to simulate the \ex{14}C-$n$ interaction.
We consider a usual Woods-Saxon form factor
\beq
V(r)=V_\ell \left[1+\exp\left(\frac{r-R_\ell}{a_\ell}\right)\right]^{-1},
\eeqn{e11}
with parameters (depth $V_\ell$, radius $R_\ell$ and diffuseness $a_\ell$)
that vary with $\ell$ to enable us to study the influence of that interaction in both the bound and continuum spectra.

 \begin{table}
 \caption{\label{t1}Parameters of the \ex{14}C-$n$ potentials used in this study.}
 \begin{ruledtabular}
 \begin{tabular}{c|c|ccc|c}
$\ell$ & label & $V_\ell$ & $R_\ell$ & $a_\ell$ &${\cal C}_{1s}$\\ 
  &         & (MeV)  & (fm)     & (fm) &  (fm$^{-1/2}$) \\\hline
$s$ & $a_s=0.6$~fm & $-52.814$ & $2.959$ & $0.6$ & 1.38\\
 and & $a_s=1.5$~fm & $-38.415$ & $2.820$ & $1.5$ & 2.29\\
 $\ell\ge3$ & $a_s=0.3$~fm & $-59.122$ & $2.959$ & $0.3$ & 1.17\\ \hline
\multirow{4}{*}{$p$} & $a_p=0.6$~fm & $-52.814$ & $2.959$ & $0.6$ & --\\
 & $E_{0p}=-8$~MeV & $-42.787$ & $2.959$ & $0.6$ & --\\
 & $a_s=1.5$~fm & $-38.415$ & $2.820$ & $1.5$ & --\\
 & $V_p=0$ & $0$ & -- & -- & --\\ \hline
$d$ & -- & $-55.993$ & $2.959$ & $0.6$ & -- \\
 \end{tabular}
 \end{ruledtabular}
 \end{table}

\begin{figure}
	\centering
	\includegraphics[width=\linewidth]{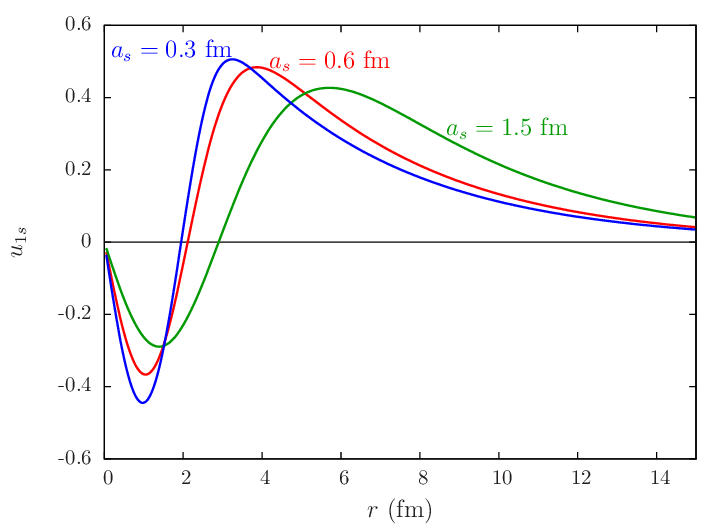}
	\caption{\label{f1} (Color online) Reduced radial ground-state wave functions of \ex{15}C obtained with the three potentials given in \tbl{t1}, labeled by their diffuseness.}
\end{figure}

In the $s$ wave, the depth of the potential is adjusted to reproduce the one-neutron separation energy $S_n(^{15}{\rm C})=1.218$~MeV, hence describing the ground state of \ex{15}C as the $1s$ state of the  Hamiltonian $H_0$ \eq{e1}.
In order to obtain different values for that bound-state ANC, we consider various geometries of the potential.
We vary mostly the diffuseness, considering first a usual value ($a_s=0.6$~fm), we then choose an unphysically large one ($a_s=1.5$~fm) in order to produce a large ANC.
We also perform our calculation with a very small diffuseness ($a_s=0.3$~fm) to obtain a small ANC.
The parameters of these potentials are listed in the upper section of \tbl{t1} together with the corresponding ANCs.
The reduced radial wave functions obtained for each of these potentials are plotted in \fig{f1}, where each curve is labelled by the diffuseness of the potential to which it corresponds.

\begin{figure}
	\centering
	\includegraphics[width=\linewidth]{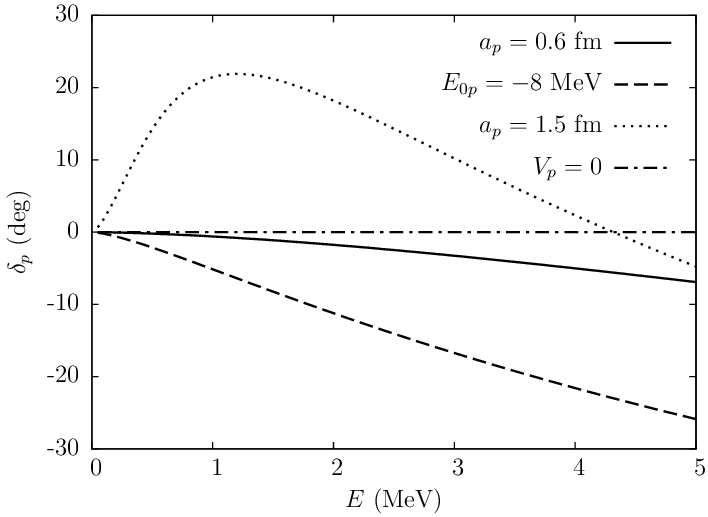}
	\caption{\label{f2} Phases shifts in the \ex{14}C-$n$ $p$ wave for the four different potentials described in \tbl{t1}.}
\end{figure}

In this model, the radiative capture proceeds mostly from a $p$ wave in the continuum towards the $1s$ bound state.
To study the influence of the description of the \ex{14}C-$n$ continuum on the reaction calculations, we also consider different potentials in the $p$ wave.
These potentials have been selected to generate significant changes in the $p$-wave phase shift $\delta_p$ (see \fig{f2}).
The parameters of these potentials are listed in the middle section of \tbl{t1}.
We first consider the same potential as in the $s$ wave with a regular diffuseness ($a_p=0.6$~fm; solid line in \fig{f2}).
Then, considering the same diffuseness, we use the prescription suggested by Summers and Nunes \cite{SN08} and fit the depth of the potential to reproduce the one-neutron separation energy of \ex{14}C in the $p$ wave ($E_{0p}=-8$~MeV; dashed line in \fig{f2}).
To fully explore the sensitivity of our calculations to the description of the continuum, we also use the very diffuse potential developed in the $s$ wave ($a_p=1.5$~fm).
That potential generates some unphysical structure in the $p$ continuum (see dotted line in \fig{f2}).
Finally, we also perform calculations with no interaction in the $p$ wave ($V_p=0$).
Accordingly, that potential generates a nil phase shift for all \ex{14}C-$n$ energies $E$ (dash-dotted line in \fig{f2}).

Our tests have shown that the reaction calculations are insensitive to the potential choice in higher partial waves ($l\ge 2$).
In the $d$ wave, we have used the same geometry as the initial potential in the $s$ wave with a small adjustment of the depth to reproduce the neutron binding energy in the $5/2^+$ excited state ($E_{0d}=-478$~keV; see last line of \tbl{t1}).
In the other partial waves ($\ell\ge3$), we use the same potential as in the $s$ wave.

In summary, this provides us with twelve different descriptions of \ex{15}C: three of the ground state combined to four of the $p$ continuum.

\subsection{Projectile-target interactions}
To describe the interaction between the projectile constituents and the target in the Coulomb-breakup calculations, we follow \Ref{CEN12} and use the Bechetti and Greenlees parametrization \cite{BG69} for the $n$-Pb optical potential.
For the $^{14}$C-Pb interaction, we use the potential developed in \Ref{Rou88} that has been adjusted to reproduce the elastic scattering of $^{16}$O on Pb at 94~MeV/nucleon.
The radius of that core-target potential is scaled by $(14^{1/3}+208^{1/3})/(16^{1/3}+208^{1/3})$ to account for the size difference between $^{14}$C and $^{16}$O.
The details of these interactions are provided in Table~II of the supplemental material of \Ref{CEN12}.

\section{\label{sensitivity}Sensitivity of the reaction calculations to the \ex{15}C description}

\subsection{\label{c15bu}Coulomb breakup of \ex{15}C at 68~MeV/nucleon}
The Coulomb breakup of \ex{15}C on Pb has been measured at RIKEN at 68~MeV/nucleon \cite{Nak09}.
In that experiment, the \ex{14}C core and the valence neutron have been detected in coincidence after dissociation and their relative-energy spectrum has been reconstructed.
The data have been selected for two ranges of the \ex{14}C-$n$ center-of-mass scattering angle: at all angles ($\Theta<6^\circ$) and at forward angles ($\Theta<2.1^\circ$).

To model this reaction, we use the Dynamical Eikonal Approximation (DEA) introduced in \Sec{DEA} and detailed in Refs.~\cite{BCG05,GBC06}.
The numerical details of the calculations are provided in \Ref{CEN12} and its supplemental material.

\begin{figure}
	\centering
	\includegraphics[width=\linewidth]{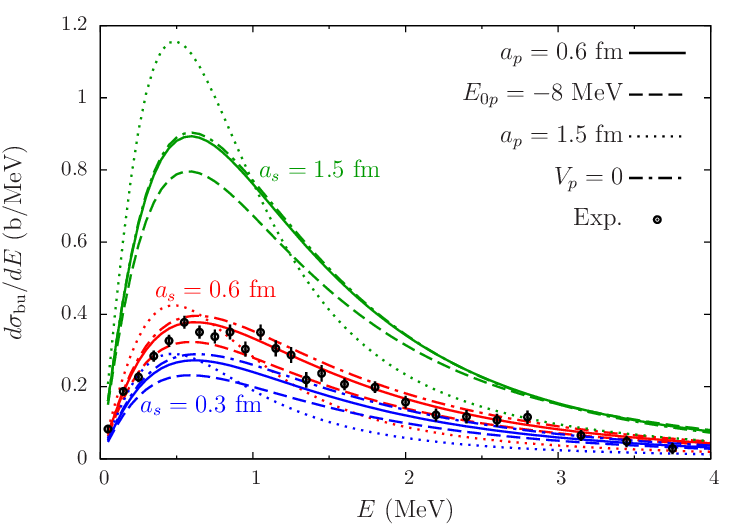}
	\caption{\label{f3} (Color online) Breakup cross section for \ex{15}C impinging on Pb at 68~MeV/nucleon as a function of the \ex{14}C-$n$ relative energy $E$.
The calculations, folded with the experimental energy resolution, have been obtained with twelve different descriptions of \ex{15}C.
The color correspond to the different bound states (see \fig{f1}) while the line types distinguish the different descriptions of the $p$ continuum (see \fig{f2}).
Experimental data are from \Ref{Nak09} and correspond to the selection over the whole experimental angular range ($\Theta<6^\circ$).
}
\end{figure}

Figure~\ref{f3} summarizes the results of our calculations.
It displays the breakup cross section as a function of the relative energy $E$ between the \ex{14}C core and the valence neutron after dissociation selected in the large experimental angular range ($\Theta<6^\circ$).
Similar results are obtained at forward angles ($\Theta<2.1^\circ$).
The DEA calculations have been performed with the twelve different descriptions of \ex{15}C detailed in \Sec{pot}, which are obtained by combining the three different potential geometries used for its initial bound state [$a_s=0.6$~fm (red lines), $a_s=1.5$~fm (green lines), and $a_s=0.3$~fm (blue lines)] and the four potentials used in the $p$ continuum [$a_p=0.6$~fm (solid lines), $E_{0p}=-8$~MeV (dashed lines), $a_p=1.5$~fm (dotted lines), and $V_p=0$ (dash-dotted lines)].

The computed cross sections spread on a large range.
They cluster into three groups, corresponding to the three descriptions of the $^{15}$C ground state.
As expected from \Ref{SN08}, they scale with ${\cal C}_{1s}^2$, the square of the ANC of that state: the cross sections obtained using the diffuse potential in the $s$ wave ($a_s=1.5$~fm, green curves), which generates the largest ANC (see \tbl{t1}), are above those obtained with the regular diffuseness ($a_s=0.6$~fm, red curves), which themselves are higher than those corresponding to the narrower potential ($a_s=0.3$~fm, blue curves).

\begin{figure}
	\centering
	\includegraphics[width=\linewidth]{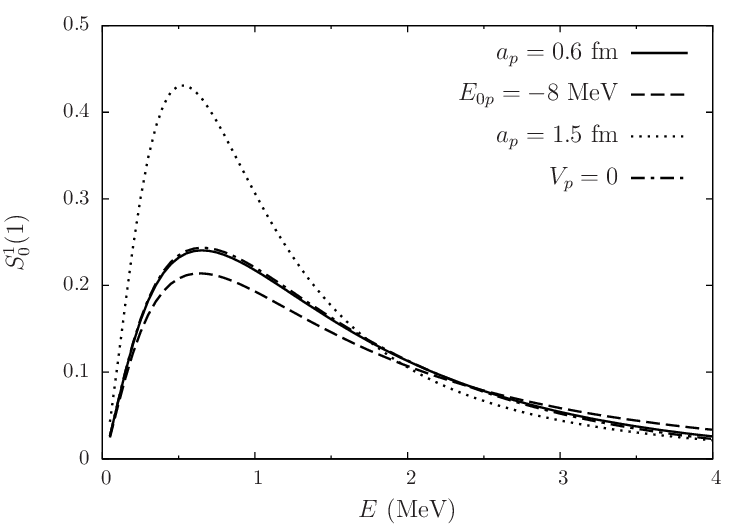}
	\caption{\label{f4} Role of the description of the \ex{14}C-$n$ continuum on breakup calculations illustrated by the dimensionless function $S_0^1(1)$ introduced by Typel and Baur \cite{TB04,TB05}.}
\end{figure}

Although they confirm the importance of the ANC in breakup calculations, these results also clearly show that this variable is not the only one at stake in this reaction.
Beside the ANC dependence, we observe that the four different descriptions of the continuum lead to different shapes of the cross sections, and these shapes are nearly independent of the value of the bound-state ANC.
In each group of curves, the solid ($a_p=0.6$~fm) and dash-dotted ($V_p=0$) lines lie very close to one another, the dashed line ($E_{0p}=-8$~MeV) is slightly below, while the dotted line ($a_p=1.5$~fm) exhibits a narrower distribution at low energy.
As suggested in \Ref{CN06}, this dependence is related to the $p$ phase shift.
This is qualitatively explained by the seminal work of Typel and Baur performed at the first order of the perturbation theory \cite{TB04,TB05}.
Assuming a purely peripheral process and a one-step $\rm E1$ transition from an $s$ bound state to the $p$ continuum, the influence of that continuum is captured in their dimensionless function $S_0^1(1)$.
That function is displayed in \fig{f4} for the four different potentials considered in the $p$ wave (see \tbl{t1}).
It qualitatively produces the dependence observed in each of the three groups of breakup cross sections obtained with our dynamical model.

This set of calculations hence confirms preliminary works, which showed that breakup calculations are mostly peripheral, in the sense that they probe only the tail of the wave function \cite{CN07,SN08,Esb09}, and that the continuum plays a significant role in these calculations \cite{TB04,TB05,CN06}.
More interestingly, it indicates that the ANC extracted with the method proposed by Summers and Nunes \cite{SN08} will be marred by the effect of the continuum, which has been overlooked in Refs.~\cite{SN08,Esb09}.
The question we will address in the next sections is how the continuum affects radiative-capture calculations and if it can be properly taken into account to extract reliable cross sections of astrophysical interest from breakup measurements.

\subsection{\label{c14ng}Radiative capture \ex{14}C$(n,\gamma)$\ex{15}C}
\begin{figure}
	\centering
	\includegraphics[width=\linewidth]{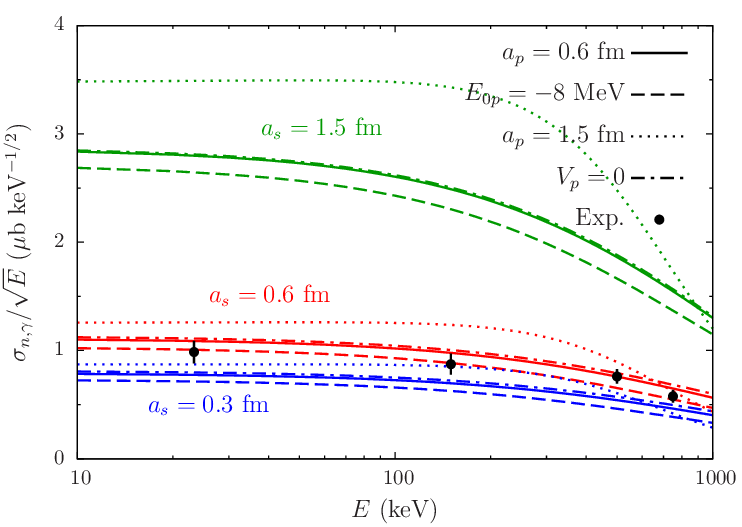}
	\caption{\label{f5a} (Color online) Cross sections for the radiative capture \ex{14}C$(n,\gamma)$\ex{15}C as a function of the \ex{14}C-$n$ relative energy $E$.
Calculations with the twelve different descriptions of \ex{15}C are shown in the same line type and color code as in \fig{f3}.
The experimental data are from \Ref{Rei08}.}
\end{figure}

In \fig{f5a}, we present the cross section $\sigma_{n,\gamma}$ computed for the radiative capture of a neutron by \ex{14}C as a function of their relative energy $E$.
We consider the twelve descriptions of \ex{15}C detailed in \Sec{pot} using the same line type and color code as in \fig{f3}.
Together with our calculations, we plot the cross sections measured by Reifart \etal\ \cite{Rei08}.
Note that we follow Summers and Nunes \cite{SN08} and scale the 23.3~keV data point by 0.67 to account for the fact that it corresponds to a Maxwellian averaged cross section (see also the Appendix of \Ref{Esb09}).

We observe the same ordering of the curves as in \fig{f3}.
They are clustered in the same three groups corresponding to the ANC of the \ex{15}C ground state.
Within each group we again observe variations that are related to the $p$ continuum description.
This shows that both the Coulomb breakup and the radiative capture depend in the same way upon the structure of \ex{15}C: they are both peripheral processes, whose cross section scales roughly with the square of the ANC of the \ex{15}C ground state and that depends on $\delta_p$ in the \ex{14}C-$n$ continuum.
Accordingly there should be a way to relate the former to the latter.

\section{\label{improvement}Relating the Coulomb breakup of \ex{15}C to the radiative capture \ex{14}C$(n,\gamma)$\ex{15}C}
\subsection{\label{scaling}Initial idea}
\begin{figure}
	\centering
	\includegraphics[width=\linewidth]{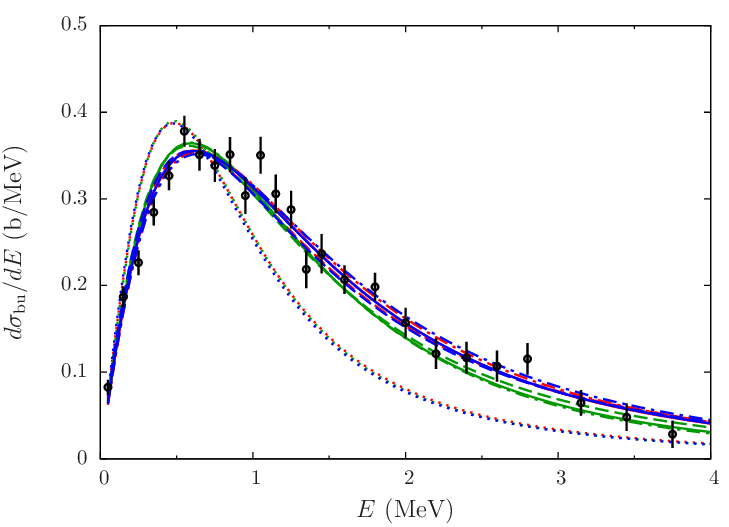}
	\caption{\label{f5} (Color online) Breakup calculations scaled to the RIKEN experimental data \cite{Nak09} ($\Theta<6^\circ$). Calculations with the twelve different descriptions of \ex{15}C are shown in the same line type and color code as in \fig{f3}.}
\end{figure}

In their original idea, Summers and Nunes suggest to extract an ``experimental'' ANC from the comparison of dynamical calculations to the RIKEN breakup data.
They then use that ANC to compute a reliable radiative-capture cross section \cite{SN08}.
As we have seen in \Sec{c15bu}, an ANC obtained in such a way will bear the trace of the continuum description.
To confirm this, we pursue the following procedure.
We scale our calculations to the RIKEN data, i.e., we multiply each of the breakup cross sections displayed in \fig{f3} by a factor chosen to minimize the $\chi^2$ to the breakup data of \Ref{Nak09}.
The resulting cross sections are plotted in \fig{f5}.
Once scaled, most of the calculations agree very well with the breakup data, as one would expect if the cross section depended solely on the ground-state ANC.
The only exceptions are the three curves obtained with the diffuse potential in the $p$ continuum (dotted lines), which exhibit too narrow a peak to reproduce the experimental energy dependence.
Albeit unphysical, this unusual geometry of the potential in the $p$ wave helps us apprehend the sensitivity of breakup calculations to the description of the continuum of the projectile.

\begin{figure}
	\centering
	\includegraphics[width=\linewidth]{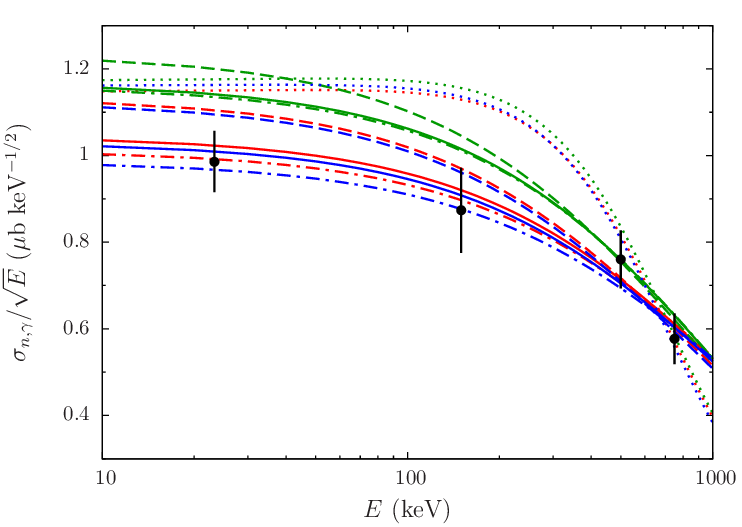}
	\caption{\label{f6} (Color online) Theoretical radiative-capture cross sections, multiplied by the scaling factor extracted from \ex{15}C Coulomb-breakup measurements, are confronted with the experimental data of Reifart \etal\  \cite{Rei08}.}
\end{figure}

The scaling factor extracted from this fit is then used to compute the radiative-capture cross section.
For this, we simply multiply the cross section provided by \Eq{e9} for each of the \ex{15}C description by its corresponding scaling factor.
If that factor depended only on the ANC, the scaled radiative-capture cross sections would be very close to each other.
Instead, we obtain the results displayed in \fig{f6}.

Although the agreement with the data of Reifart \etal\ \cite{Rei08} is not bad, we observe a much larger spread in this way of extracting the radiative-capture cross section from Coulomb breakup than obtained by Summers and Nunes \cite{SN08}.
Moreover the average value at low energy overestimates the direct measurements.
Most of the problem arises from the cross sections obtained with the very diffuse potential, both in the \ex{15}C ground state (green curves) and in the $p$ continuum (dotted lines).
However, since both reactions are sensitive to the same aspects of the description of the nucleus, we look for a way to improve the ANC method by properly taking that description into account.

\subsection{\label{lowE}Selecting data at low energy}
Interestingly, all breakup distributions exhibit very similar behaviors at low energy: they all rise steeply with $E$ (see \fig{f5}) and the slope at the origin varies with the description of the $p$ continuum.
This can be qualitatively understood from the low-energy expansion of the $S^1_0$ function provided by Typel and Baur (see Eq.~(10) of \Ref{TB04}), which depends on the scattering length in the $p$ continuum.
This suggests that a more effective scaling factor could be extracted from the data if the fit were performed at low energy, e.g., below $E=0.5$~MeV.
In this way, the scaling would naturally absorb the low-energy description of the continuum.
It would also make more sense on a physics viewpoint because these low energies are closer to the range at which the radiative-capture cross sections are needed for astrophysical purposes.
In this way, this scaling method is not affected by the description of the continuum at higher energies, which has no influence at astrophysical energies.

\begin{figure}
	\centering
	\includegraphics[width=\linewidth]{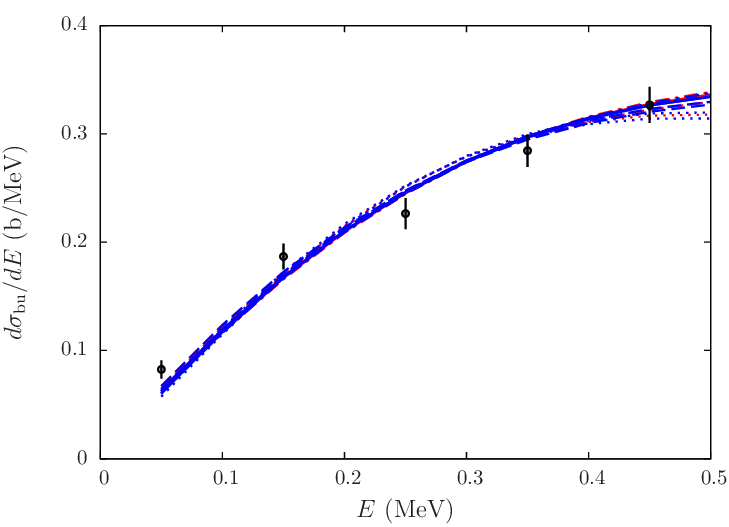}
	\caption{\label{f7} (Color online) Theoretical breakup cross sections for $^{15}$C scaled to the data of \Ref{Nak09} limited to $E=0.5$~MeV in the \ex{14}C-$n$ continuum ($\Theta<6^\circ$).}
\end{figure}

\begin{figure}
	\centering
	\includegraphics[width=\linewidth]{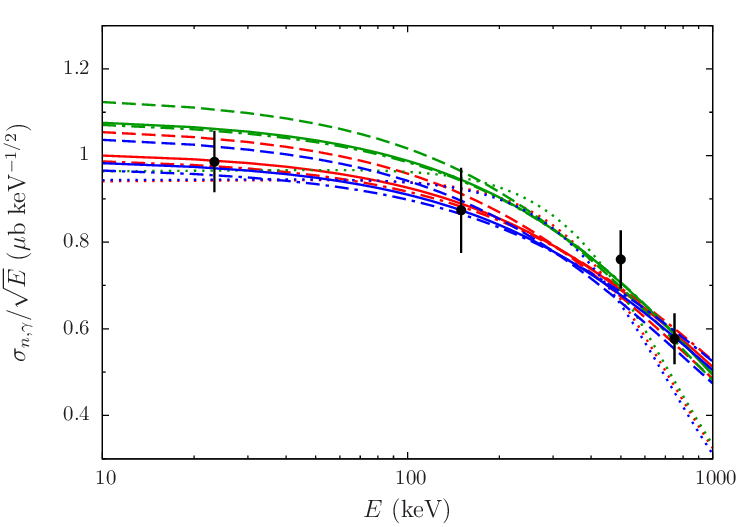}
	\caption{\label{f8} (Color online) Theoretical radiative-capture cross sections multiplied by the scaling factor extracted from \ex{15}C Coulomb breakup limited to 0.5~MeV in the \ex{14}C-$n$ continuum.}
\end{figure}

The suggested scaling is illustrated in \fig{f7}.
With that focus on the low energy, the spread between the different calculations is strongly reduced compared to \fig{f5}.
Even the calculations performed with the very diffuse potential in the $p$ continuum (dotted lines) are nearly superimposed on the other scaled cross sections.
Considering the scaling factors extracted from this fit, we repeat the radiative-capture calculations and obtain the cross sections displayed in \fig{f8}.
Compared to \fig{f6}, the spread is significantly reduced, even in the cases in which 
unrealistic potentials are used.
Selecting data at low energy is thus not only more meaningful in a physics viewpoint, it also naturally reduces the theoretical uncertainty due to the description of the projectile continuum.

\subsection{\label{fwd}Selecting data at forward angle}
Beside the large scattering-angle range ($\Theta <6^{\circ}$) considered in the two previous sections, Nakamura \etal\ have also measured the Coulomb breakup of \ex{15}C selecting the data at forward angle ($\Theta <2.1^{\circ}$) \cite{Nak09}.
At these smaller angles, the influence of the nuclear interaction between the projectile and the target is strongly reduced \cite{GCB07}.
Such an angular cut also reduces the higher-order effects in the breakup, like couplings within the continuum \cite{CB05}.
This set of data seems therefore better suited to extract radiative-capture cross sections from breakup measurements.

We have repeated the scaling procedure described in Secs.~\ref{scaling} and \ref{lowE} using the set of data limited to the forward angles.
When the fit is performed on the whole energy range, i.e., up to 4~MeV, we obtain a noticeable reduction of the spread of the predicted radiative-capture cross section compared to the large angular range.
This reduction is similar to what has been obtained when the data are selected at low energy (see \fig{f8}).

\begin{figure}
	\centering
	\includegraphics[width=\linewidth]{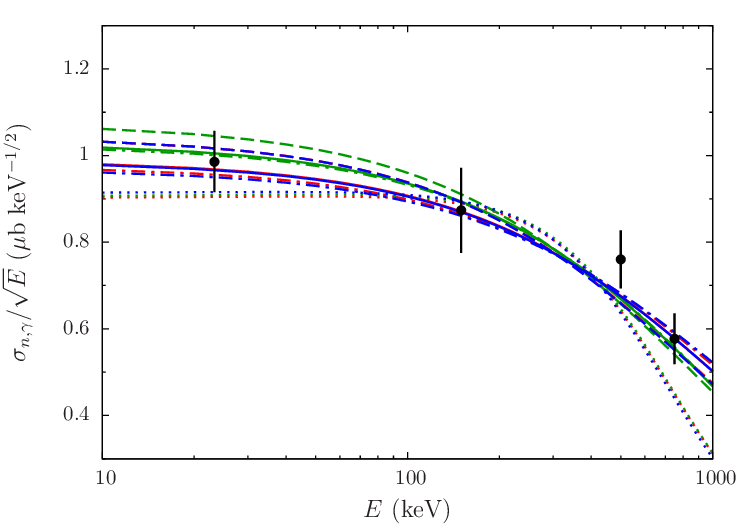}
	\caption{\label{f9} (Color online) Theoretical radiative-capture cross sections multiplied by the scaling factor extracted from \ex{15}C Coulomb breakup selected at forward angles ($\Theta <2.1^{\circ}$) and limited to $E=0.5$~MeV in the \ex{14}C-$n$ continuum.}
\end{figure}

The major improvement comes from the combination of both ideas: fitting the breakup calculations only at low energy and for data selected at forward angles (see \fig{f9}).
First, the spread in our predictions is significantly reduced to the point that it is now similar to that of the direct measurements.
Second, the average value of the radiative-capture cross sections extracted from the breakup data are in perfect agreement with the Reifart \etal\ data.
Interestingly, these excellent results are obtained for all the descriptions of \ex{15}C, including the most exotic ones.

At the lowest energy ($E=23.3$~keV) our estimate of the radiative-capture cross section obtained from the different descriptions of \ex{15}C and restricting the fit of the breakup data to low energy and forward angle is $4.74\pm0.27~\mu$b, which is in excellent agreement with the direct measurement: $4.76\pm0.34~\mu$b (see \Ref{Rei08} with the correction mentioned in Refs.~\cite{SN08,Esb09}).
The uncertainty on this estimate would be significantly reduced if the calculations performed with the unrealistic descriptions of \ex{15}C---viz. those involving the very diffuse potential in the bound state (green lines) or in the continuum (dotted lines)---were ignored.
On a physics viewpoint, it means that if some constraint can be put on the description of the continuum, e.g. via an estimate of the scattering length from experiment or a reliable microscopic calculation, the uncertainty of the present method would be strongly reduced.
Nevertheless, our analysis shows that, even without any information on the continuum description of the projectile, selecting the breakup data to low core-neutron energy and forward scattering angle enables a clean and reliable extraction of radiative-capture cross sections from Coulomb-breakup measurements.

\section{\label{conclusion}Conclusion}
Coulomb breakup has been proposed as an indirect technique to infer radiative-capture cross sections of astrophysical interest \cite{BBR86,BR96,BHT03}.
The main reasoning behind this idea is that---at least at the first order of the perturbation theory---the former can be seen as the time-reversed reaction of the latter.
Unfortunately, later analyses have shown that higher-order effects in breakup reactions, such as couplings within the continuum, are non-negligible and that they should be accounted for in order to describe correctly the reaction process \cite{CB05,EBS05,SN08,Esb09}.

Basing their idea on the fact that breakup \cite{CN07} and radiative capture \cite{TBD06} are mostly peripheral, Summers and Nunes have proposed to use Coulomb-breakup measurements to extract an ``experimental'' ANC for the bound state of the nucleus.
That ANC can then be used to compute a good estimate of the radiative-capture cross section \cite{SN08}.
This result was confirmed by Esbensen \cite{Esb09}.
In the present work, we analyse the sensitivity of this new method to the description of the continuum of the nucleus, which also affects breakup calculations \cite{CN06}.

As in Refs.~\cite{SN08,Esb09}, we focus on $^{15}$C, for which both the Coulomb breakup \cite{Nak09} and the radiative capture $^{14}$C$(n,\gamma)$$^{15}$C \cite{Rei08} have been measured accurately.
Our analysis confirms the peripherality of the breakup reaction \cite{CN07} and its sensitivity to the core-neutron continuum \cite{CN06}.
Although the latter effect hinders the extraction of an exact ANC, we have shown that selecting the data at low core-neutron energy and forward scattering angles enables us to improve the method suggested by Summers and Nunes \cite{SN08}.
The former condition enables us to reliably absorb the description of the core-neutron continuum.
It also corresponds to choosing a meaningful energy range, where radiative-capture cross sections are needed for astrophysical applications.
The latter condition reduces the effects of the nuclear interaction between the projectile and the target, and the higher orders, such as couplings within the continuum.
The scaling factor extracted in this way from the $\chi^2$ fit of dynamical calculations to breakup data leads to a small spread of the radiative-capture calculations and an excellent agreement with direct measurements.

In the particular case of the radiative-capture $^{14}$C$(n,\gamma)$, our method provides a cross section of $4.74\pm0.27~\mu$b at $E=23.3$~keV from the RIKEN breakup experiment \cite{Nak09}.
This value is in full agreement with the direct measurement of Reifart \etal\ \cite{Rei08}.

This new method hence revives the Coulomb breakup technique to infer radiative-capture cross sections at astrophysical energies.
Unlike the method proposed by Summers and Nunes, the scaling we suggest accounts for the description of the projectile continuum.
This enables us to significantly reduce the uncertainty in the deduced radiative-capture cross sections.
Our study thus demonstrates that experimental works should focus on the low-energy and forward-angle ranges, where data provide the best indirect predictions of radiative-capture cross sections of astrophysical applications and, thanks to the method exposed here, in a nearly model-independent way.

In the future, it would be interesting to study the extension of this method to proton captures and see in particular if it can resolve the discrepancy that remains between the $^7$Be$(p,\gamma)$$^8$B cross sections measured directly and those extracted from the Coulomb breakup of $^8$B \cite{AGR11}.



\appendix
\section{Erratum}

The excellent agreement observed above was obtained following the prescription of Summers and Nunes to rescale the low-energy data point, assuming it corresponded to a Maxwellian average on the neutron energy \cite{SN08}.
Unfortunately, as pointed out by Esbensen and Reifarth \cite{Esb09err}, this is not the case.
To be properly compared to the capture measurements, calculations should be averaged over the energy distribution of the neutron beams used in the experiment \cite{FR17p} (see Fig.~3 of \Ref{Rei08}).

 \begin{table*}
 \begin{ruledtabular}
 \begin{tabular}{cc|cccccccc}
\multicolumn{2}{c|}{} &  \multicolumn{8}{c}{$\sigma_{n,\gamma}$ ($\mu$b)} \\
\multicolumn{2}{c|}{Potential} & \multicolumn{2}{c}{$E=23.3$~keV} & \multicolumn{2}{c}{$E=150$~keV} & \multicolumn{2}{c}{$E=500$~keV} & \multicolumn{2}{c}{$E=750$~keV}\\
$s$ wave & $p$ wave & Unscaled & Scaled & Unscaled & Scaled & Unscaled & Scaled & Unscaled & Scaled\\ \hline
\multirow{4}{*}{$a_s=0.6$~fm} & $a_p=0.6$~fm & 6.55   & 5.41 & 11.85 & 9.79 & 16.61 & 13.72 & 17.50 & 14.45\\
 & $E_{0p}=-8$~MeV & 6.03   & 5.64 & 10.70 & 10.01 & 14.38 & 13.46 & 14.87 & 13.91\\
 & $a_p=1.5$~fm & 7.74   & 5.06 & 14.95 & 9.78 & 20.13 & 13.16 & 18.07 & 11.82\\
 & $V_p=0$ & 6.69   & 5.35 & 12.16 &  9.72  & 17.26 & 13.80 & 18.36 & 14.68\\ \hline
\multirow{4}{*}{$a_s=1.5$~fm} & $a_p=0.6$~fm & 16.81 & 5.59 & 30.15 & 10.02 & 40.98 & 13.61 & 42.13 & 14.00\\
 & $E_{0p}=-8$~MeV & 15.82 & 5.78 & 27.91 & 10.20 & 36.76 & 13.43 & 37.38 & 13.66\\
 & $a_p=1.5$~fm & 22.45 & 5.08 & 41.48 & 9.83 & 55.92 & 13.25 & 50.17 & 11.89\\
 & $V_p=0$ & 16.89 & 5.57 & 30.33 & 10.00 & 41.39 & 13.64 & 42.62 & 14.04\\ \hline
\multirow{4}{*}{$a_s=0.3$~fm} & $a_p=0.6$~fm & 4.66   & 5.40 & 8.44  & 9.77 & 11.84 & 13.70 & 12.49 & 14.45\\
 & $E_{0p}=-8$~MeV & 4.27 & 5.64 & 7.58   & 10.01 & 10.17 & 13.43 & 10.50 & 13.86\\
 & $a_p=1.5$~fm & 5.36   & 5.12 & 10.31 & 9.84 & 13.69 & 13.08 & 12.16 & 11.62\\
 & $V_p=0$ & 4.81 & 5.32 & 8.75 &  9.68 & 12.48 & 13.81 & 13.33 & 14.75\\ \hline
\multicolumn{2}{r|}{Average $\pm$ standard deviation} & $10\pm 6$ & $5.41\pm0.23$ & $18\pm11$ & $9.89\pm0.15$ & $24\pm15$ & $13.51\pm0.24$ & $24\pm14$ & $13.6\pm1.1$\\
\multicolumn{2}{r|}{Contribution from the $5/2^+$ state} & & 0.26 & & 0.52 & & 0.98 & & 1.25\\
\multicolumn{2}{r|}{Total} & & $5.67\pm0.23$ & & $10.41\pm0.15$ & & $14.49\pm0.24$ & & $14.8\pm1.1$\\
\multicolumn{2}{r|}{Experiment \cite{Rei08}} & & $7.1\pm0.5$ & & $10.7\pm1.2$ & & $17.0\pm1.5$ & &$15.8\pm1.6$\\
 \end{tabular}
  \caption{\label{tE1}Theoretical radiative-capture cross sections obtained with the twelve $^{14}$C-$n$ potentials of \Ref{CN17} averaged over the energy distribution of the neutron beam used in the experiment of \Ref{Rei08}.
The measured values are listed in the last row.}
 \end{ruledtabular}
 \end{table*}

In this Erratum, we perform this averaging in a systematic way for the twelve $^{14}$C-$n$ potentials developed \Sec{pot}.
The results are summarized in \tbl{tE1}.
Each line of that table corresponds to one potential identified by its geometry in the $s$ partial wave of the $^{15}$C ground state and in the $p$ wave of the $^{14}$C-$n$ continuum (see \Sec{pot}).
The numbers in the various columns provide the radiative-capture cross section $\sigma_{n,\gamma}$ obtained from averaging the theoretical prediction over the energy distribution of the neutron beams used in the experiment of Reifarth \etal\ \cite{Rei08}.
For each of the beam energies considered ($E=23.3$, 150, 500, and 750~keV) we provide the value obtained with the bare potentials (denoted ``Unscaled'') and the ``Scaled'' cross sections resulting from the multiplication of the unscaled result by the factor obtained from the fit of our breakup calculation with the data of Nakamura \etal\ \cite{Nak09}.
That scaling is performed at low energy ($E<0.5$~MeV) and forward scattering angle ($\Theta<2.1^\circ$), which was found to be the most accurate one, as it focuses on the energy range of astrophysical interest and the angles at which the nuclear interaction between the projectile and the target is the less significant, i.e. where the process is fully dominated by the Coulomb interaction.
The fourth line before last provides the average and standard deviation over the twelve potentials.
As already mentioned above, we see that once scaled all potentials provide nearly identical radiative-capture cross sections, despite displaying very different ``unscaled'' cross sections.
Accordingly, the standard deviation of the ``scaled'' results is tremendously reduced compared to that obtained for the ``unscaled results''.

To properly confront these estimates to the experimental data (last line of \tbl{tE1}), we add to that main contribution the cross section for the capture to the $5/2^+$ bound excited state of $^{15}$C, which we have neglected before.
We describe this state as a $0d5/2$ neutron bound to the $^{14}$C ground state and consider the same twelve $^{14}$C-$n$ potentials as in \Sec{pot}.
Similarly to the capture to the ground state, the dominant sensitivity comes from this excited-state ANC, the description of the $p$ continuum leads to only 5\% uncertainty.
The estimate provided in the third to last line of \tbl{tE1} corresponds to the most usual geometry of the potential ($a_p=a_s=0.6$~fm).
As already seen in Refs.~\cite{Rei08,Esb09}, that contribution amounts to a mere 5\% of the total cross section; a rough estimate is therefore sufficient here.

The total of these two contributions is shown in the penultimate line of \tbl{tE1}.
Thanks to the scaling suggested in our study, this total is close to the experimental value.
However, contrary to the results discussed above, we observe a systematic underestimation of the experiment by the theory prediction, similar to the one obtained by Esbensen \cite{Esb09}.
Although our scaling method enables us to account for the influence of both the ANC and the continuum description, and hence reduce the uncertainty in extracting the radiative-capture cross section from Coulomb breakup measurements, it does not fully reconcile both methods.

\begin{acknowledgments}
We thank D.~Baye for interesting discussions on this work and for providing us with the code that computes the radiative-capture cross sections.
We thank C.~Forss\`en and R.~Reifarth for interesting discussions on this averaging problem and for providing us with the numerical estimate of the neutron energy distributions.
This work is part of the Belgian Research Initiative on eXotic
nuclei (BriX), program P7/12 on inter-university attraction
poles of the Belgian Federal Science Policy Office.
It was supported in part by the Research Credit No. 19526092 of the Belgian Funds for Scientific Research F.R.S.-FNRS.
This project has received funding from the European Union's Horizon 2020 research and innovation programme under grant agreement No 654002.
P.~C. acknowledges the support of the Deutsche Forschungsgesellschaft (DFG) with the Collaborative Research Center 1245.
\end{acknowledgments}



\end{document}